\documentclass[aps,showpacs,prl,toolkits,twocolumn]{revtex4}
\usepackage{epsfig,amssymb,amsfonts,amsmath,pstricks,psfrag,graphicx}

\makeatletter
\renewcommand{\@makefntext}[1]{\parindent=1em\noindent\hbox to 1.8em
{\hss$^{\@thefnmark}$}#1}
\renewcommand{\@footnotemark}{\hbox{\mathsurround=0pt$^{\@thefnmark}$}}

\makeatother

\begin{document}
\title{Second order chiral restoration phase transition at low
temperatures in quarkyonic matter}
\author{ L. Ya. Glozman and R. F. Wagenbrunn}
\affiliation{Institute for
 Physics, Theoretical Physics branch, University of Graz, Universit\"atsplatz 5,
A-8010 Graz, Austria}

\newcommand{\be}{\begin{equation}}
\newcommand{\bea}{\begin{eqnarray}}
\newcommand{\ee}{\end{equation}}
\newcommand{\eea}{\end{eqnarray}}
\newcommand{\ds}{\displaystyle}
\newcommand{\low}[1]{\raisebox{-1mm}{$#1$}}
\newcommand{\loww}[1]{\raisebox{-1.5mm}{$#1$}}
\newcommand{\lmn}{\mathop{\sim}\limits_{n\gg 1}}
\newcommand{\vpint}{\int\makebox[0mm][r]{\bf --\hspace*{0.13cm}}}
\newcommand{\too}{\mathop{\to}\limits_{N_C\to\infty}}
\newcommand{\vp}{\varphi}
\newcommand{\vx}{{\vec x}}
\newcommand{\vy}{{\vec y}}
\newcommand{\vz}{{\vec z}}
\newcommand{\vk}{{\vec k}}
\newcommand{\vq}{{\vec q}}
\newcommand{\vpp}{{\vec p}}
\newcommand{\vn}{{\vec n}}
\newcommand{\vg}{{\vec \gamma}}

\begin{abstract}
In this Addendum to our recent paper, Phys. Rev. D 77, 054027 (2008),
we point out that a chiral restoration phase transition in a quarkyonic
matter at low temperatures is of  second order within a manifestly
confining and chirally symmetric large $N_c$ model. This result is 
qualitatively different as compared to  NJL and NJL-like models that
are not confining
and might have some implications for the existence or nonexistence of 
the critical end point in the QCD phase diagram.
\end{abstract}

\maketitle

In a recent paper \cite{GW} we studied a chiral restoration phase
transition at finite density and zero temperature within the only
known exactly solvable manifestly chirally $SU(2)_L \times SU(2)_R$
symmetric and confining 
model in four dimensions \cite{GNJL}. It is assumed within this large $N_c$
model that
the only gluonic interaction is a linear confining potential of the Coulomb
type. Then the chiral symmetry breaking can be obtained from the 
Schwinger-Dyson (gap) equation, while the color-singlet mesons are
solutions of the Bethe-Salpeter equation. A single-quark Dirac operator
is always infrared-divergent and hence a single quark is confined. 
All color-singlet observable
quantities  (quark condensate, hadron mass, ...)
are infrared-finite and well defined. 
We have previously applied this model to study  the chiral restoration 
in excited mesons \cite{WG}.

A decisive feature of this model is that even above the chiral restoration 
point
at a critical chemical potential the system is still in a confining
mode and the only possible excitations are chirally invariant color-singlet
hadrons (or meson-like color-singlet particle-hole excitations) \cite{GW,G}.
This is because in the large $N_c$ limit there are no vacuum quark
loops and no Debye screening of the confining
gluon propagator at any chemical potential.
The masses of the color-singlet and chirally-invariant hadrons
increase with the chemical potential. This model represents a possible
microscopic scenario for a recently proposed quarkyonic matter \cite{pisarski}.

In this Addendum to our paper \cite{GW} we would like to point out
 that the chiral restoration phase transition at low temperatures is
 of  second order. In order to see it explicitly we show in  Fig. 1
 the dependence of the chiral angle, $\varphi_p$, which is a solution
 of a gap equation, on momentum $p$ as well as on the quark Fermi-momentum
 $p_f$ in a system with a finite chemical potential. The chiral restoration
 phase transition happens at the Fermi-momentum 
 $p_f^{cr} = 0.109 \sqrt \sigma$,
 where $\sigma$ is the string tension that supplies a dimensional scale
 in our task. At $p_f = p_f^{cr}$ the chiral angle takes its trivial
 value, $\varphi_p = 0$, and the chiral symmetry gets restored. An
 important feature is that the chiral angle approaches this trivial
 value  {\it continuously}. 
\begin{figure}
\includegraphics[width=0.8\hsize,clip=]{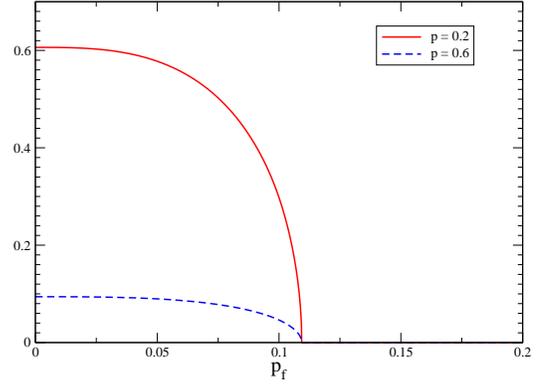}
\caption{Chiral angle $\varphi_p$
as a function of the Fermi momentum $p_f$ for two given values $p$. 
Both momenta are units of $\sqrt \sigma$.}
\end{figure}
\begin{figure}
\includegraphics[width=0.8\hsize,clip=]{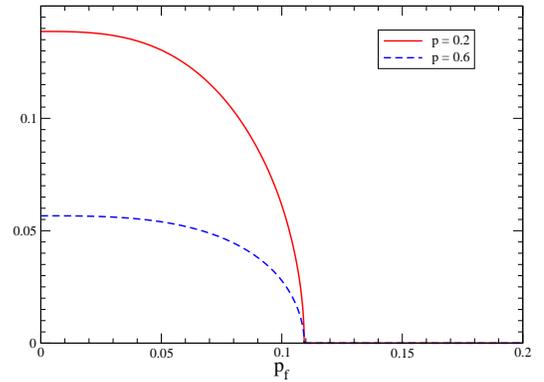}
\caption{Dynamical mass of quarks $M(p)$
as a function of the Fermi momentum $p_f$ for two given values of $p$. 
Both momenta are units of $\sqrt \sigma$.}
\end{figure}

 The chiral angle uniquely defines a dynamical mass $M(p)$ of quarks,
 that is associated with the chiral symmetry breaking, as well as a
 quark condensate.  Fig. 2 illustrates the {\it continuous} approach of
 the dynamical mass of quarks to its trivial value $M(p)=0$ at the
 chiral restoration point. These two figures clearly demonstrate
 that the chiral restoration phase transition in the chiral limit
 is of  second order.
 The same can actually be seen from the continuous behavior of the quark
 condensate as well as the meson masses at the chiral restoration point
 (see Figs. 1 and 6 of Ref. \cite{GW}).
 
 It is an interesting feature of this chirally symmetric
 and manifestly confining model. Indeed, it is well known that
 the Nambu and Jona-Lasinio models (NJL) exhibit a first order
 chiral restoration phase transition at low temperatures \cite{NJL}.
 The same is true within the Polyakov-improved NJL models (PNJL) \cite{PNJL},
 that simulate confinement in a statistical manner via a coupling
 of the nonconfining NJL Hamiltonian with the Polyakov loop. It has been 
 suggested very
 recently that this model also exhibits a chirally symmetric and confining 
 quarkyonic phase \cite{PNJLF,PNJLA}. The chiral restoration phase transition
 is, however, of  first order, like in the standard NJL model. This
 is in contrast to our manifestly confining model. One can speculate that
 this feature of the PNJL models is related to the absence of an explicit
 confinement of quarks.
 
 Our results might have some implications for the existence/nonexistence
 of the critical end point, that is a subject of  significant 
 experimental and  theoretical interest. The chiral restoration
 phase transition at zero density and large temperature is of 
 second order in the chiral limit and becomes a crossover with realistic
 quark masses \cite{Karsch,Fodor}. Then, if at low temperatures and large 
 density a
 chiral restoration phase transition is of  first order, as suggested
 by the NJL-like and some other nonconfining models, there
 must exist a critical end point in the QCD phase diagram. If, however,
 a chiral restoration phase transition at low temperatures is of 
 second order or a crossover, then there should be no critical end
 point (or, as an exotic possibility, there might be two critical
 end points).
 
 Our results have been obtained within a large $N_c$ model, where
 the vacuum fermion loops as well as the Debye screening of the confining
 potential are absent. It would be interesting to see how persistent
 our results might be beyond the large $N_c$ limit.
 
 Another caveat is that for the ground state of the system the implicit
 assumption is used that this ground state can be approximated as a quark Fermi
 gas, like in the NJL model. However, within the confining model the
 ground state is unlikely to be a pure Fermi gas of quarks. Going beyond
 a simple Fermi gas description of the ground state is a complicated
 task and should  be considered as an important direction in the future.

\medskip
{\bf Acknowledgements}
L.Ya.G. is thankful to T. Hatsuda, M. Huang, D. Kharzeev,
L. McLerran, 
V. Miransky, R. Pisarski, B.-J. Schaefer,  E. Shuryak, D. Son, J. Verbaarshot 
and J. Wambach for discussions and 
acknowledges support of the Austrian Science
Fund through the grant P19168-N16.


\begin{thebibliography}{99}



\bibitem{GW} L. Ya. Glozman and R. F. Wagenbrunn, Phys. Rev. D {\bf 77},
054027 (2008). 
\bibitem{GNJL}  A. Le Yaouanc, L. Oliver, O. Pene, and J. C.Raynal,
Phys. Rev. D {\bf 29}, 1233  (1984); {\bf 31}, 137  (1985);
S.~L. Adler and A.~C. Davis, Nucl. Phys. B {\bf 244},  469 (1984);
A. Kocic, Phys. Rev. D {\bf 33}, 1785  (1986);
R. Alkofer and P.~A. Amundsen, Nucl. Phys. B {\bf 306}, 305  (1988);
P. Bicudo and J. E. Ribeiro,
Phys. Rev. D {\bf 42}  (1990) 1611; {\bf 42}, 1625 (1990);
P. J. A. Bicudo and A. V. Nefediev, 
Phys. Rev. D {\bf 68}, 065021 (2003);
 F. J. Llanes-Estrada and S. R. Cotanch, Phys. Rev. Lett.,
{\bf 84}, 1102 (2000);
R. Alkofer, M. Kloker, A. Krassnigg, R. F. Wagenbrunn,
 Phys. Rev. Lett., {\bf 96}, 022001 (2006).
\bibitem{WG} R. F. Wagenbrunn and L. Ya. Glozman,
Phys. Lett. {\bf B 643}, 98 (2006); R. F. Wagenbrunn and L. Ya. Glozman,
Phys. Rev. {\bf D 75}, 036007 (2007);
L. Ya. Glozman, Phys. Rep. {\bf 444},1 (2007).
\bibitem{pisarski} L. McLerran and R. D. Pisarski, Nucl. Phys. A {\bf 796},
83 (2007).
\bibitem{G} L. Ya. Glozman, arXiv:0803.1636 [hep-ph] 
\bibitem{NJL} M. Asakawa and K. Yazaki, Nucl. Phys. A {\bf 504}, 668 (1089);
B. Barducci et al, Phys. Lett. B {\bf 231}, 463 (1989); Phys. Rev. D {\bf 41},
1610 (1990); S. P. Klevansky, Rev. Mod. Phys. {\bf 64}, 646 (1992).
\bibitem{PNJL} P. N. Meisinger and M. C. Ogilvie, Phys. Lett. B {\bf 379}, 163
(1996); K. Fukushima, Phys. Lett. B {\bf 591}, 277 (2004); C. Ratti,
M. A. Thaler, W. Weise, Phys. Rev. D {\bf 73}, 014019 (2006); 
E. Megias, E. Ruiz Arriola, L.L. Salcedo,  Phys. Rev. D {\bf 74}, 114014 (2006);
S. K. Ghosh et al,  Phys. Rev. D {\bf 73} 114007 (2006);
B. J. Schaefer, J. M. Pawlowski and J. Wambach,  Phys. Rev. D {\bf 76}, 
074023 (2007);
C. Sasaki, B. Friman and K. Redlich, Phys. Rev. D {\bf 75}, 074013 (2007). 
\bibitem{PNJLF} K. Fukushima, arXiv:0803.3318 [hep-ph]
\bibitem{PNJLA} H. Abuki, R. Anglani, R. Gatto, G. Nardulli, M. Ruggieri,
arXiv:0805.1509 [hep-ph]
\bibitem{Karsch} F. Karsch, Plenary talk at Lattice 2007, PoS
LATTICE2007:015,2006
\bibitem{Fodor} Z. Fodor et al, Plenary talk at Lattice 2007; 
LATTICE2007:189,2006

\end{thebibliography}
\end{document}